\documentclass[preprint,aps,showpacs]{revtex4}
\usepackage{graphicx}
\usepackage{epstopdf}
\usepackage{dcolumn}
\usepackage{bm}

\begin{document}


\title{Electronic structure and magnetic properties of the graphene/Fe/Ni(111) intercalation-like system}

\author{M.\,Weser,$^1$ E.\,N.\,Voloshina,$^2$ K.\,Horn,$^1$ and Yu.\,S.\,Dedkov$^{1,}$\footnote{Corresponding author. E-mail: dedkov@fhi-berlin.mpg.de}}
\affiliation{$^1$Fritz-Haber Institut der Max-Planck Gesellschaft, 14195 Berlin, Germany}
\affiliation{$^2$Institut f\"ur Chemie und Biochemie-Physikalische und Theoretische Chemie, Freie Universit\"at Berlin, Takustra\ss e 3, 14195 Berlin, Germany}

\date{\today}

\begin{abstract}
The electronic structure and magnetic properties of the graphene/Fe/Ni(111) system were investigated via combination of the density functional theory calculations and electron-spectroscopy methods. This system was prepared via intercalation of thin Fe layer (1\,ML) underneath graphene on Ni(111) and its inert properties were verified by means of photoelectron spectroscopy. Intercalation of iron in the space between graphene and Ni(111) changes drastically the magnetic response from the graphene layer that is explained by the formation of the highly spin-polarized $3d_{z^2}$ quantum-well state in the thin iron layer.  
\end{abstract}


\maketitle

Carbon is a widespread element in nature: it is found in abundance in the sun, stars, comets, and atmospheres of most planets. Atomic carbon is a very short-lived species and, therefore, carbon is stabilized in various multi-atomic structures (carbon allotropic forms) depending on the hybridization of carbon atoms: $sp^3$ hybridized carbon atoms form diamond, $sp^2$ hybridization of carbon atoms leads to the formation of two-dimensional (2D) flat sheets consisting of hexagonal rings and loosely bonded through weak van der Waals forces forming bulk graphite, mixed $sp^2/sp^3$ hybridization leads to the arrangements of carbon atoms in fullerenes or carbon nanotubes.

For a long time, the magnetism in the carbon-based systems was under debate (for review, see Refs.~\cite{Fan:2010,Yazyev:2010}). Usually, the magnetism was considered only for the systems containing unpaired $d$- or $f$-electrons, neglecting the correlations between $s,p$-electrons that lead to the long-range magnetic ordering in materials that do not contain $d,f$-electrons in the non-fully filled electronic shells. Experimentally, the magnetic state in the carbon-based systems (proton-irradiated graphite) was recently found~\cite{Esquinazi:2003,Ohldag:2007}, where magnetism was attributed to point defects~\cite{Duplock:2004,Yazyev:2007}. In case of the reduced dimensionality of the zigzag-edged graphene nanoflakes the large net spin moment was predicted~\cite{Fernandez-rossier:2007,Wang:2008}.

The exceptional transport properties of graphene, a single atom layer of hexagonally coordinated carbon ($sp^2$-hybridized), make it a promising material for applications in microelectronics and sensing~\cite{Geim:2007a,Neto:2009}. Among the exciting properties of graphene are a long electronic mean free path~\cite{Geim:2007a} and negligible spin-orbit coupling in graphene~\cite{Min:2006} leading to large spin relaxation times, which render this material ideal for ballistic spin transport. Graphene-based spin electronic devices possess a tremendous potential for high-density non-volatile memories, reconfigurable electronic devices and, possibly, solid-state quantum computing elements~\cite{Falko:2007,Rycerz:2007,Trauzettel:2007}. One of the potential application of graphene is use of it as a junction layer in spin-filtering devices~\cite{Karpan:2007,Karpan:2008}. Such devices will require new materials to overcome some of the major problems currently hindering progress, such as low spin injection efficiency~\cite{Schmidt:2000,Wolf:2001}.

Recent demonstration of spin injection in graphene~\cite{Tombros:2007} opens a new road in application of this 2D material in spintronics questioning electronic, magnetic, and interfacial properties of the graphene/ferromagnet layered system, which have to be studied in details. In our latest works we have demonstrated by means of x-ray magnetic circular dichroism (XMCD) and spin-resolved photoelectron spectroscopy that the net magnetic moment of about $0.05-0.1\mu_B$ per carbon atom is induced in the graphene layer via its contact with ferromagnetic Ni(111) substrate~\cite{Weser:2010,Dedkov:2010a}. The magnetic properties of the graphene layer (net magnetic moment as well as exchange interaction) in this system are expected to be improved via intercalation of thin Fe layers underneath graphene on Ni(111) due to the larger magnetic moment of an Fe atom. Moreover, the magnetoresistance ratio for the FM/graphene/FM sandwich is changed from 16\% to 61\% when FM=Ni(111) is replaced by Fe(111) layer~\cite{Yazyev:2009}. 

Here we present complex studies of the electronic and magnetic properties of the graphene/Fe/Ni(111) intercalation-like system by means of photoelectron spectroscopy of core levels as well of valence band, x-ray absorption spectroscopy (XAS), and XMCD at the Ni, Fe $L_{2,3}$ and C $K$ absorption edges. The C $1s\rightarrow\pi^*,\sigma^*$ XMCD spectra reveals an induced magnetic moment of the carbon atoms in the graphene layer aligned parallel to the Ni $3d$ and Fe $3d$ magnetization. It is found that intercalation of Fe between graphene and Ni(111) changes drastically the magnetic response from the graphene layer. Obtained experimental results are compared with the density-functional theory calculations and magnetic moment of carbon atoms in the graphene layer is estimated before and after Fe intercalation in the graphene/Ni(111) system.

The experimental results were obtained at the D1011 beamline of the MAX-lab (Lund, Sweden). The procedure of the sample preparation and experimental conditions are identical to the one described in Refs.~\cite{Dedkov:2008a,Dedkov:2008b,Dedkov:2008,Weser:2010,Dedkov:2010a}. The quality, homogeneity, and cleanliness of the prepared systems were verified by means of low-energy electron diffraction (LEED) and core-level as well as valence-band photoemission. XAS spectra were collected at Ni, Fe $L_{2,3}$ and C $K$ absorption edges in partial (repulsive potential $U=-100$\,V) and total electron yield modes (PEY and TEY, respectively) with an energy resolution of 80\,meV. Magnetic dichroism spectra were obtained with circularly polarized light (degree of polarization is $P=0.75$) in the remanence magnetic state of the system after applying of an external magnetic field of 500\,Oe along the $<1\bar{1}0>$ easy magnetization axis of the Ni(111) film. All experiments were performed at 300\,K.

In our DFT studies, the electronic and structural properties of the graphene-substrate system are obtained using the Perdew-Burke-Ernzerhof (PBE) functional~\cite{Perdew:1996}. For solving the resulting Kohn-Sham equation we have used the Vienna Ab Initio Simulation Package (VASP)~\cite{Kresse:1996a,Kresse:1996b} with the projector augmented wave basis sets~\cite{Blochl:1994}. The plane-wave kinetic energy cut-off is set to 500\,eV. The supercell used to model the graphene-metal interface is constructed from a slab of $13$ layers of metal atoms with a graphene sheet adsorbed at both sides and a vacuum region of approximately $14$\,\AA. In optimizing the geometry, the positions ($z$-coordinates) of the carbon atoms as well as those of the top two layers of metal atoms are allowed to relax. In the total energy calculations and during the structural relaxations the $k$-meshes for sampling the supercell Brillouin zone are chosen to be as dense as $24\times24$ and $12\times12$, respectively. The detailed analysis of theoretical results is presented as supplementary material.

\begin{figure}[t]
\includegraphics[scale=2.0]{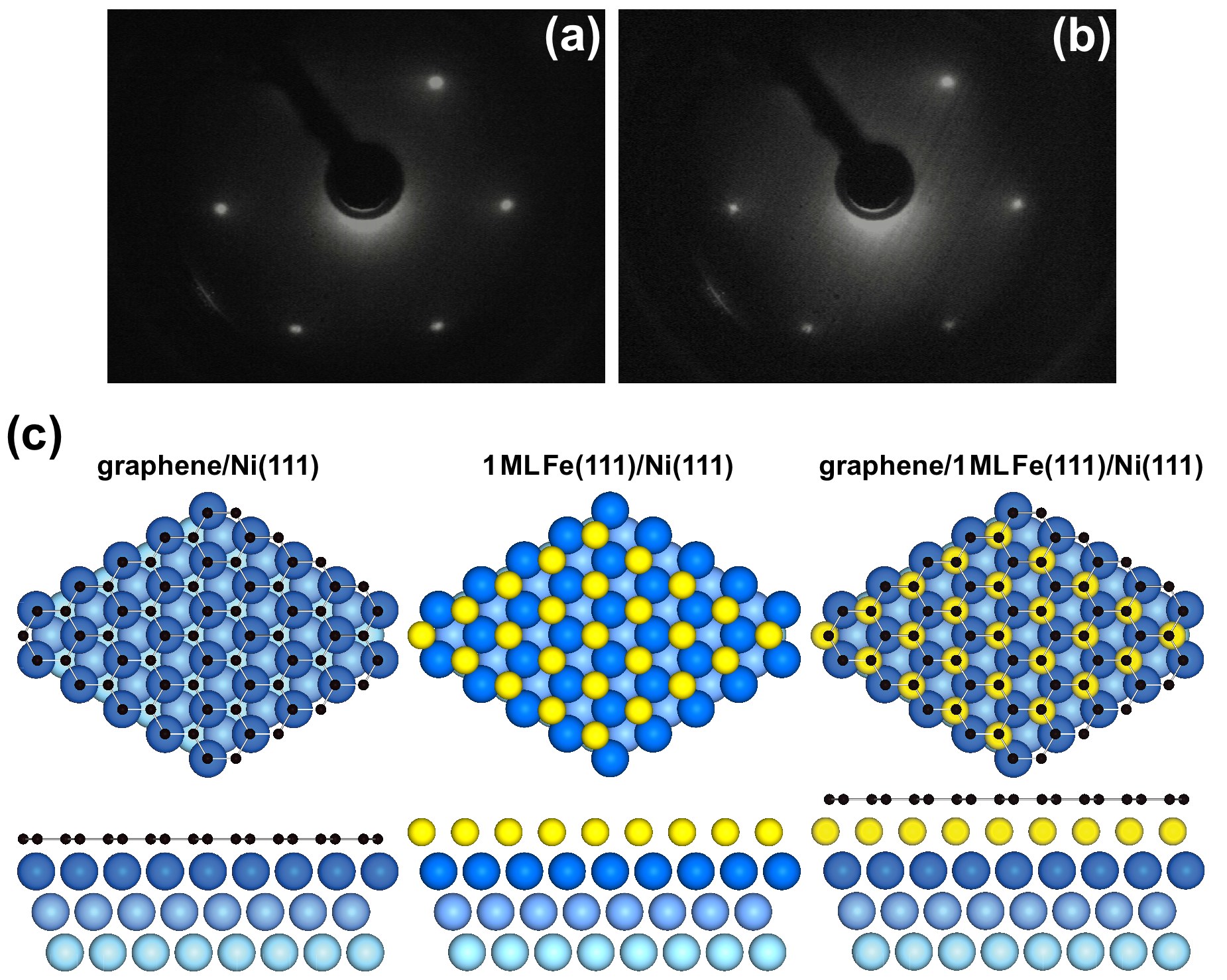}
\caption{(Color online) LEED images of the (a) graphene/Ni(111) and the (b) graphene/1\,ML\,Fe(111)/Ni(111) systems collected at 120\,eV and 125\,eV of the primary electron energy, respectively. (c) Top- (upper panel) and side-views (lower panel) of graphene/Ni(111), 1\,ML\,Fe(111)/Ni(111), and graphene/1\,ML\,Fe(111)/Ni(111). For discussion of the structures see text.}
\label{LEEDandSTR}
\end{figure}

Fig.~\ref{LEEDandSTR} shows the LEED images of (a) graphene/Ni(111) and (b) graphene/1\,ML\,Fe(111)/Ni(111). The widely accepted structure of graphene/Ni(111) is when carbon atoms are arranged in the so-called $top-fcc$ configuration on Ni(111)~\cite{Gamo:1997,Bertoni:2004,Karpan:2007} [Fig.~\ref{LEEDandSTR}(c, left)]. Our calculations also confirm this model. After intercalation of Fe underneath graphene on Ni(111) the three-fold symmetry of the system is preserved as deduced from the LEED analysis. In the consideration of the possible crystallographic structures of the intercalation-like system, Fe atoms below graphene layer can be placed either in the $fcc$ or in the $hcp$ hollow sites above the Ni(111) slab. According to the symmetry of the system obtained after Fe intercalation (also confirmed by DFT calculations discussed later) the two most energetically favorable configurations of the graphene layer and the iron atoms in the graphene/Fe/Ni(111) system are: (i) Fe atom is placed in the $hcp$ hollow site and carbon atoms are in the $top-hcp$ configuration with respect to Ni(111); (ii) Fe atom is placed in the $fcc$ hollow site and carbon atoms are in the $top-fcc$ configuration with respect to Ni(111). The later arrangement is shown in Fig.~\ref{LEEDandSTR}(c, right) and it presents the case of $TOP-HCP$ configuration of graphene layer on metallic surface where one of the carbon atoms from the graphene unit cell is placed above interface metal atom (Fe) and the second one is in the $HCP$ position with respect to the metal stack (avove Ni).

\begin{figure}[t]
\begin{center}
\includegraphics[scale=0.375]{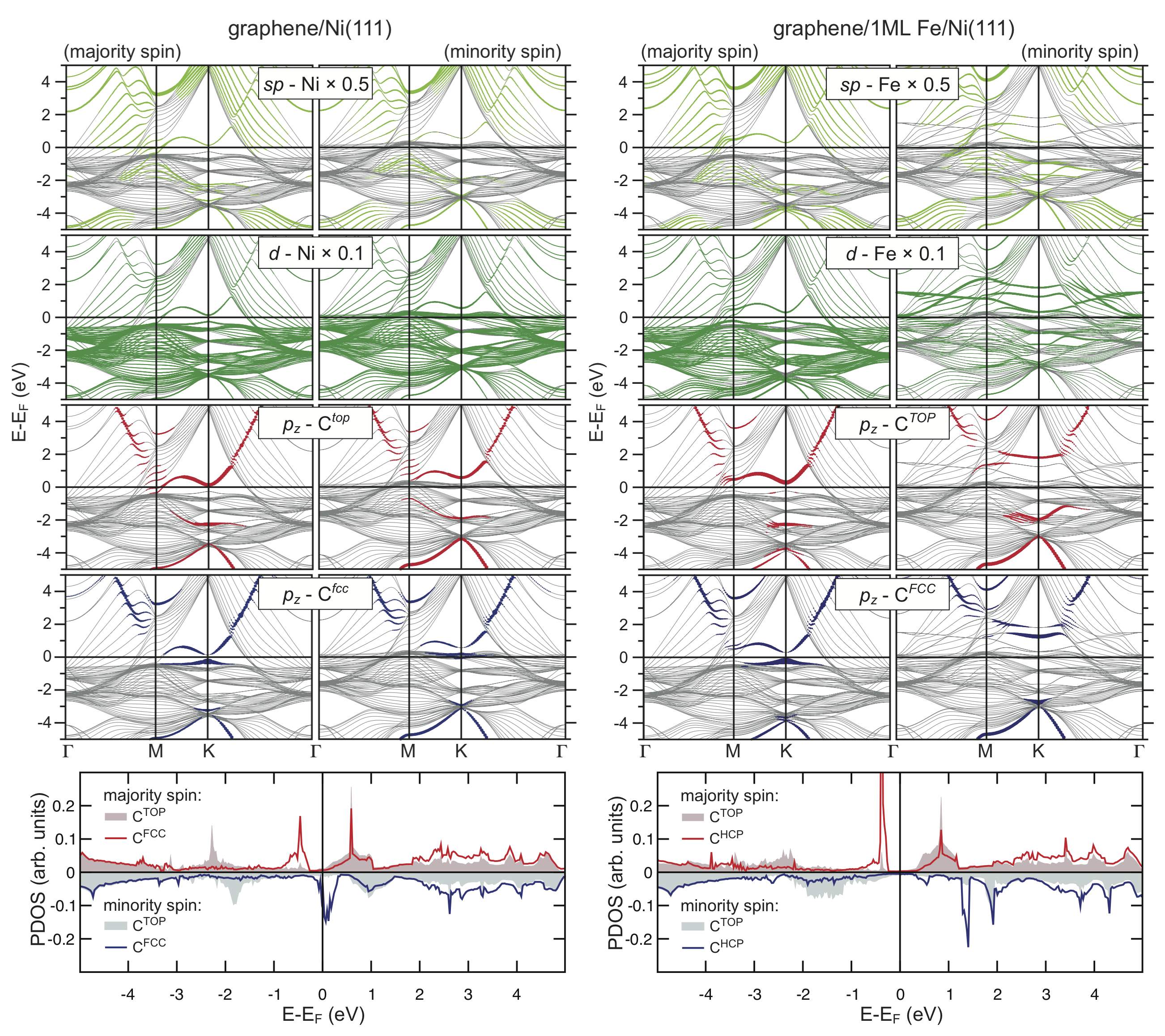}
\end{center}
\caption{(Color online) Electronic band structure together with the partial band characters (upper rows) and the corresponding C-atom projected density of states (lower row) for the graphene/Ni(111) and graphene/1\,ML\,Fe(111)/Ni(111) systems.}
\label{bandstrPDOS}
\end{figure}

The comparison of the calculated spin-resolved band structures of graphene/Ni(111) and graphene/1\,ML\,Fe(111)/Ni(111) is presented in Figs.~\ref{bandstrPDOS} (see also supplementary material). Our results for the calculated electronic and magnetic structure of the graphene/Ni(111) interface are in very good agreement with previous calculations~\cite{Bertoni:2004,Karpan:2007,Karpan:2008} (Fig.~\ref{bandstrPDOS}, left panel): the $top-fcc$ arrangement of the carbon atoms on Ni(111) is the most energetically favorable with the mean distance between graphene and Ni(111) of $2.134$\,\AA; magnetic moment of the interface Ni atoms is $-0.025\,\mu_B$ and $0.543\,\mu_B$ for $4sp$ and $3d$ valence electrons, respectively, compared to $-0.03\,\mu_B$ and $0.71\,\mu_B$ for the clean Ni(111) surface; magnetic moments of C-$top$ and C-$fcc$ are $-0.019\,\mu_B$ and $0.031\,\mu_B$, respectively; the assignment of the electronic states in the valence band around the Fermi level ($E_F$) can be performed according to Ref.~\cite{Bertoni:2004} (see also supplementary material).

For the graphene/1\,ML\,Fe(111)/Ni(111) system the most energetically favorable configuration is when Fe atoms are placed in the $fcc$ hollow sites and carbon atoms are in the $top-fcc$ configuration with respect to Ni(111) [Fig.~\ref{LEEDandSTR} (c)]. In this case the positions of carbon atoms with respect to the underlying metal layers are $TOP-HCP$ and is different compared to the one for graphene/Ni(111). The interaction between the graphene layer and underlying Fe is stronger that is reflected in the shorter graphene-Fe interface distance: atoms C-$top$ and C-$hcp$ are placed by $2.114$\,\AA\ and $2.089$\,\AA\ above Fe layer, respectively. Deposition of graphene on 1\,ML\,Fe(111)/Ni(111) leads to the decreasing of the magnetic moment of interface Fe: $-0.035\,\mu_B$ and $2.469\,\mu_B$ for $4sp$ and $3d$ valence electrons of Fe, compared to $-0.028\,\mu_B$ and $2.622\,\mu_B$ in the 1\,ML\,Fe(111)/Ni(111) system. The induced magnetic moment of carbon atoms in the graphene layer for this system is stronger compared to graphene/Ni(111): $-0.050\,\mu_B$ and $0.039\,\mu_B$ for C-$TOP$ and C-$HCP$, respectively. This increasing can be assigned to the larger magnetic moment of the underlying Fe atoms compared to Ni.

The electronic band structure of the graphene/1\,ML\,Fe(111)/Ni(111) system (Fig.~\ref{bandstrPDOS}, right panel) is modified in comparison with the one for graphene/Ni(111). In contrast to the majority-spin band structure, which is similar for both systems (except some small energy shifts of the graphene $\pi$ states and interface states), the minority-spin band structures are different. Initially, for the 1\,ML\,Fe(111)/Ni(111) system, there is a one quantum-well state at $1.47$\,eV above $E_F$ of the minority-spin Fe $3d_{z^2}$ character in the energy gap around the $K$ point (see supplementary material). After adsorption of graphene, this state is split in two states according to the existence of the two inequivalent carbon atoms placed on the different adsorption positions, $TOP$ and $HCP$, respectively. The minority-spin components of these two interface states, $I_3$ and $I_4$ (according to notation of Bertoni \textit{et al.}~\cite{Bertoni:2004}), are shifted further upwards above $E_F$, compared to graphene/Ni(111), and can be found at the $K$ point at $1.19$\,eV and $1.72$\,eV, respectively. These two states are the result of hybridization of the C $p_z$ orbitals of graphene with the quantum-well state of the $3d$ character of the underlying Fe layer. The state at $1.72$\,eV originates from the hybridization of the $p_z$ orbital of the C-$TOP$ atom and the $3d_{z^2}$ orbital of the interface Fe atom. The state at $1.19$\,eV is the result of hybridization of the $p_z$ orbital of the C-$HCP$ atom and mainly the $3d_{xz},3d_{yz}$ orbitals of the interface Fe atom (Fig.~\ref{bandstrPDOS}, right panel; see also supplementary material for complete analysis). The corresponding differences between electronic structures of graphene/Ni(111) and graphene/1\,ML\,Fe(111)/Ni(111) are also reflected in the partial density of states for graphene layer shown in Fig.~\ref{bandstrPDOS} (lower row).

The magnetic properties of the graphene layer on the ferromagnetic Ni(111) surface were studied in details in Refs.~\cite{Weser:2010,Dedkov:2010a} and it was shown that strong hybridization of the graphene $\pi$ and Ni $3d$ states leads to the appearance of the induced magnetic moment of carbon atoms with a value of $0.05-0.1\,\mu_B$ per carbon atom. The respective reduction of the magnetic moment of the Ni interface atoms, compared to the bulk value, was predicted and observed in the former experiments.

\begin{figure}[t]
\includegraphics[scale=0.375]{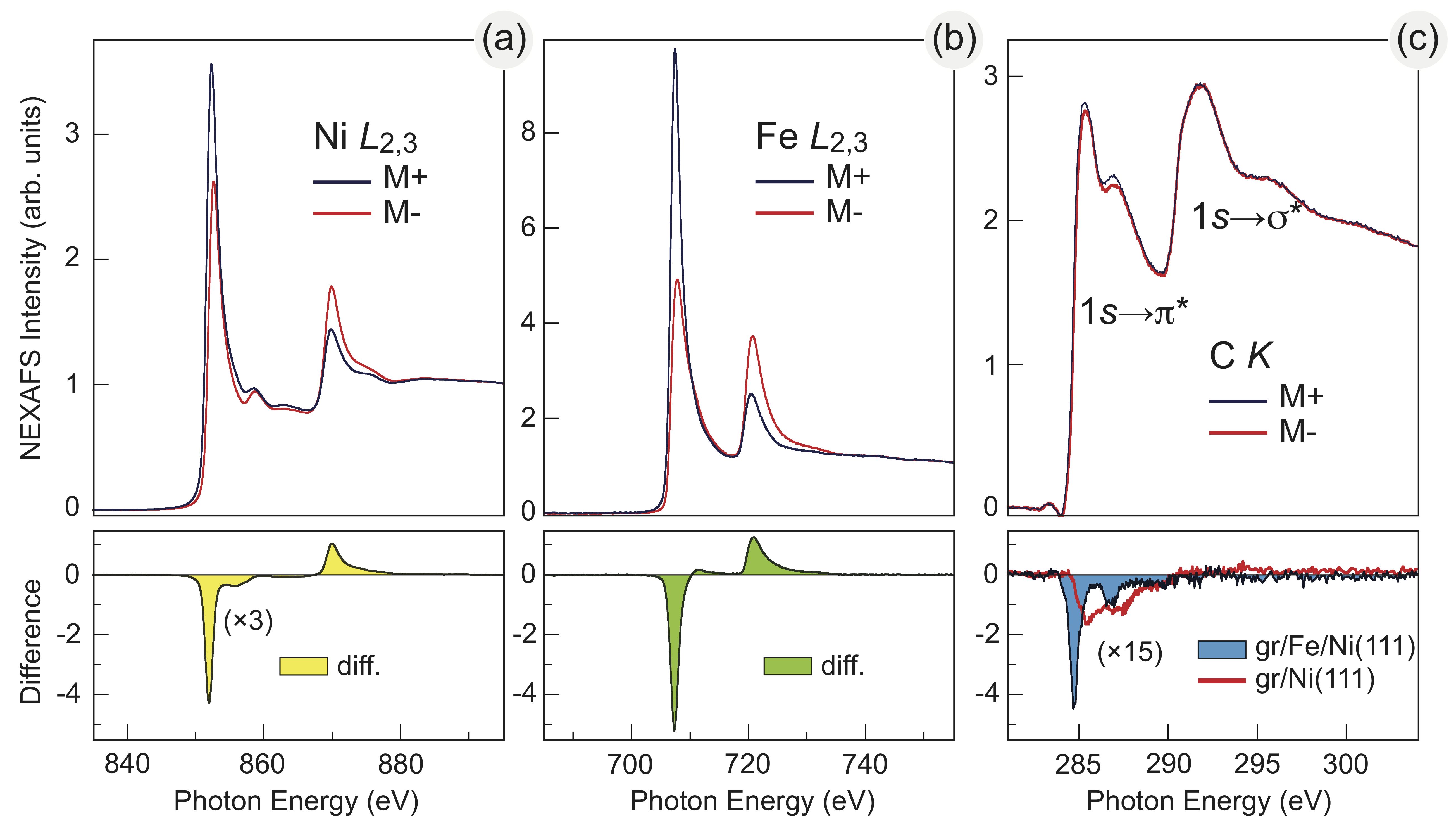}
\caption{(Color online) XMCD spectra of the graphene/1\,ML\,Fe(111)/Ni(111) system measured at the (a) Ni $L_{2,3}$, (b) Fe $L_{2,3}$, and (c) C $K$ absorption edges. The upper panels show the absorption spectra measured with the circularly polarized light for two opposite magnetization directions and the lower panels present the corresponding difference (in case of Ni $L_{2,3}$ and C $K$ edges the differences are multiplied by factors 4 and 15, respectively). The XMCD difference obtained for the graphene/Ni(111) system (multiplied by factor 15) is reproduced from Ref.~\cite{Dedkov:2010a} in the lower panel of (c).}
\label{XMCD}
\end{figure}

The results of the investigation of the magnetic properties of the graphene/1\,ML\,Fe(111)/Ni(111) system are summarized in Fig.~\ref{XMCD} where the absorption spectra measured with the circularly polarized light for two opposite magnetization directions of the sample and the corresponding XMCD difference are shown in the upper and lower panel, respectively, for the (a) Ni $L_{2,3}$, (b) Fe $L_{2,3}$, and (c) C $K$ absorption thresholds. The XMCD spectrum collected at the Ni $L_{2,3}$ edge in the TEY mode is in perfect agreement with the previously published data~\cite{Srivastava:1998,Dhesi:1999,Nesvizhskii:2000}. The bulk values of the spin- and orbital-magnetic moments $\mu_S=0.69\,\mu_B$ and $\mu_L=0.07\,\mu_B$ of Ni calculated from the spectra on the basis of sum-rules are in very good agreement with previously published experimental values~\cite{Srivastava:1998,Baberschke:1996} as well as with the spin-magnetic moment $\mu_S=0.67\,\mu_B$ calculated for the graphene/Ni(111) system (see Ref.~\cite{Bertoni:2004} and supplimentary material). The Fe $L_{2,3}$ XMCD spectrum is in good agreement with previously presented data for the $fcc$ Fe~\cite{Cros:2000,Yamamoto:2010}. The spin- and orbital-magnetic moments can be estimated from these data assuming the number of holes in the iron layer in the graphene/1\,ML\,Fe(111)/Ni(111) system to be $n_h=3.7$ (for bulk Fe $n_h=3.4$; transfer of $0.15e^-$ from Fe to Ni~\cite{Wu:1992} and the same value from Fe to graphene layer~\cite{Yamamoto:1992}; the present theoretical calculations give value of $n_h=3.691$ for Fe $3d$ states). This number leads to $\mu_S=2.56\pm0.1\,\mu_B$ and $\mu_L=0.31\pm0.05\,\mu_B$ for the spin- and orbital-magnetic moments of Fe atoms in the intercalated layer which are in very good agreement with value of spin-magnetic moment $2.469\,\mu_B$ for Fe obtained in calculations for graphene/1\,ML\,Fe(111)/Ni(111). The relatively large uncertainty in the value of magnetic moment extracted from experiment arises mainly from the estimation of the number of Fe $3d$ holes and from the error for the degree of circular polarization of light.

The C $K$-edge XMCD spectrum of graphene/1\,ML\,Fe(111)/Ni(111) is strongly modified compared to the one measured for the graphene/Ni(111) system~\cite{Weser:2010,Dedkov:2010a}. The most important observation is the increasing of the magnetic contrast at the C $K$ edge by factor of $\approx2.7$ which correlates with the theoretical predictions about spin-magnetic moment of carbon atoms in the graphene layer on the Ni(111) and 1\,ML\,Fe(111)/Ni(111) substrates. As in the previous case~\cite{Weser:2010} the relatively strong XMCD contrast is detected for C $1s\rightarrow\pi^*$ transitions whereas there is almost no variation of the absorption signal upon magnetization reversal is visible for the C $1s\rightarrow\sigma^*$ transitions. These observations are explained well by the theoretically predicted strong hybridization of the out-of-plane graphene $\pi$ and Fe $3d$ states and the existence of the relatively weak hybridization between in-plane graphene $\sigma$ and Fe $3d$ states, similar to the graphene on Ni(111)~\cite{Weser:2010,Dedkov:2010a}.

The C $K$-edge XAS spectrum of the graphene/1\,ML\,Fe(111)/Ni(111) system in the energy range corresponding to the $1s\rightarrow\pi^*$ transition  consists of two peaks which can be assigned, similar to Refs.~\cite{Weser:2010,Dedkov:2010a}, to transition of $1s$ electron on the interface states which are result of the hybridization of C $p_z$ orbitals of graphene layer and Fe $3d$ orbitals (see Fig.~\ref{bandstrPDOS}, right panel and supplementary material for the identification of the interface states). The modification and the increasing of the XMCD contrast going from graphene/Ni(111) to graphene/1\,ML\,Fe(111)/Ni(111) can be explained by the larger energy splitting between spin-up and spin-down C-projected density of states [Fig.~\ref{bandstrPDOS} (lower row)]. XMCD spectra measured at the C $K$-edge can only provide information on the orbital moment. From the negative sign of the XMCD signal one can conclude that the averaged orbital moment of carbon atoms of the graphene layer is aligned parallel to both, the spin and orbital moments of the substrate layer. It is noteworthy that the orientation of individual spin and orbital moments of both Fe and C at different sites cannot be determined from the experimental XMCD data. On the basis of comparison of the dichroic signals measured at the C $K$-edge for the graphene/Ni(111) and graphene/1\,ML\,Fe(111)/Ni(111) systems one can estimate the increasing of the spin-magnetic moment of carbon atoms up to $\approx0.2-0.25\,\mu_B$. Our calculations give values of magnetic moments of $-0.050\,\mu_B$ and $0.039\,\mu_B$ for C-$TOP$ and C-$HCP$, respectively. However, the magnetic splitting of the spin-up and spin-down parts of the interface states $I_3$ and $I_4$ in the graphene/1\,ML\,Fe(111)/Ni(111) system was found to be about $1.45$\,eV which may yield higher values for the magnetic moment of carbon atoms.

\textit{In conclusion}, we present studies of the electronic and magnetic properties of the graphene/Fe/Ni(111) intercalation-like system by means of photoelectron spectroscopy of core levels as well of valence band, XAS, and XMCD at the Ni, Fe $L_{2,3}$ and C $K$ absorption edges. The C $1s\rightarrow\pi^*,\sigma^*$ XMCD spectra reveals an induced magnetic moment of the carbon atoms ($\mu_S\approx0.2-0.25\,\mu_B$) in the graphene layer aligned parallel to the Ni $3d$ and Fe $3d$ magnetization. It is found that intercalation of Fe between graphene and Ni(111) changes drastically the magnetic response from the graphene layer. Obtained experimental results are compared with the density-functional theory calculations and magnetic moment of carbon atoms in the graphene layer is estimated before and after Fe intercalation in the graphene/Ni(111) system.

\textit{Note added after submission.} During preparation of the present manuscript the Ref.~\cite{Sun:2010} was published where analysis of the possible atom arrangements in graphene/Fe/Ni(111) and calculations of its electronic structure (density of states) are presented.

This work has been supported by the European Science Foundation (ESF) under the EUROCORES Programme EuroGRAPHENE (Project ``SpinGraph''). Y.\,S.\,D. acknowledges the financial support by the German Research Foundation (DFG) under project DE\,1679/2-1. E.\,N.\,V. appreciate the support from DFG through the Collaborative Research Center (SFB) 765 ``Multivalency as chemical organisation and action principle: New architectures, functions and applications''. Y.\,S.\,D and M.\,W. acknowledge the financial support and technical assistance by MAX-lab. We appreciate the support from the HLRN (High Performance Computing Network of Northern Germany) in Berlin.
\newpage
\noindent
Supplementary material for manuscript:\\
\textbf{Electronic structure and magnetic properties of the graphene/Fe/Ni(111) intercalation-like system}\\
\newline
M.\,Weser,$^1$ E.\,N.\,Voloshina,$^2$ K.\,Horn,$^1$ and Yu.\,S.\,Dedkov$^{1}$\\
\newline
$^1$Fritz-Haber Institut der Max-Planck Gesellschaft, 14195 Berlin, Germany\\
$^2$Institut f\"ur Chemie und Biochemie-Physikalische und Theoretische Chemie, Freie Universit\"at Berlin, Takustra\ss e 3, 14195 Berlin, Germany
\newline
\newline
\newline
\textbf{List of tables:}
\\
Table\,1: Parameters for the calculated atomic structures of graphene/Ni(111) and graphene/1\,ML\,Fe(111)/Ni(111).
\\
Table\,2: Binding energies of the interface states for stable crystallographic structures of graphene/Ni(111) and graphene/1\,ML\,Fe(111)/Ni(111).
\newline
\newline
\textbf{List of figures:}
\\
Fig.\,S1: Electronic structures of graphene/Ni(111) and graphene/1\,ML\,Fe(111)/Ni(111) in the wide energy range ($E-E_F: -21 ... +5$\,eV). 
\\
Fig.\,S2: Electronic structures of graphene/Ni(111) around $E_F$ with the corresponding weights of Ni- and C-projected bands.
\\
Fig.\,S3: Electronic structures of 1\,ML\,Fe(111)/Ni(111) [for crystallographic structure see Fig.\,1(c)] around $E_F$ with the corresponding weights of Ni-projected bands.
\\
Fig.\,S4: Electronic structures of 1\,ML\,Fe(111)/Ni(111) [for crystallographic structure see Fig.\,1(c)] around $E_F$ with the corresponding weights of Fe-projected bands.
\\
Fig.\,S5: Electronic structures of graphene/1\,ML\,Fe(111)/Ni(111) around $E_F$ with the corresponding weights of Fe- and C-projected bands.

\begin{table*}
\caption{\label{table1} Results for the atomic structure of the three graphene/FM interface models and for the clean FM surface: $d_0$ is the distance between the graphene overlayer and the interface FM layer (the two values for the two nonequivalent carbon atoms are indicated); $d_1$ is the distance between the interface FM layer and the second FM layer; $d_2$ is the distance between the second and third FM layers; $\Delta E$ is the energy difference between the energy calculated for the different slabs and the energy calculated for the most stable geometry; $m_{\rm FM}$ is the interface/surface FM spin magnetic moment (the two values for the $sp$ and $d$ magnetizations are indicated); $m_{\rm C}$ is the interface carbon spin magnetic moment (the two values for the two nonequivalent carbon atoms are indicated).}
\begin{ruledtabular}
\begin{tabular}{c c c c c } 
            &FM          &\multicolumn{3}{c}{Structure of the graphene/FM interface}\\
\cline{2-5}
                                             & Ni(111)                  & $top$-$fcc$             & $top$-$hcp$            &$fcc$-$hcp$\\ 
\hline
$d_0$ (\AA)                       &                                 &$2.135$/$2.133$ &$2.145$/$2.146$ &$3.540$/$3.540$\\
$d_1$ (\AA)                       &$2.005$                  &$2.020$                 &$2.020$                 &$2.020$               \\
$d_2$ (\AA)                       &$2.031$                  &$2.017$                 &$2.015$                 &$2.036$               \\
$\Delta E$ (eV)                 &                                 &$0.000$                 &$0.049$                 &$0.054$               \\
$m_{\rm FM}$ ($\mu_B$)&$-0.030$/$0.710$&$-0.025$/$0.543$&$-0.026$/$0.514$&$-0.029$/$0.669$\\
$m_{\rm C}$ ($\mu_B$)   &                                &$-0.019$/$0.031$&$-0.019$/$0.027$&$\,\,\,\,\,\,\,0.000$/$0.000$ \\
\hline
                                              & 1ML Fe/Ni(111) & $TOP$-$FCC$             & $TOP$-$HCP$            &$FCC$-$HCP$\\ 
\hline
$d_0$ (\AA)                        &                                &$2.117$/$2.092$ &$2.114$/$2.089$ &$3.487$/$3.487$\\
$d_1$ (\AA)                        &$2.039$                 &$2.029$                 &$2.019$                 &$2.028$               \\
$d_2$ (\AA)                        &$2.061$                 &$2.044$                 &$2.057$                 &$2.062$               \\
$\Delta E$ (eV)                  &                                &$0.021$                 &$0.000$                 &$0.104$               \\
$m_{\rm FM}$ ($\mu_B$)&$-0.028$/$2.622$&$-0.034$/$2.486$&$-0.035$/$2.469$&$-0.022$/$2.616$\\
$m_{\rm C}$ ($\mu_B$)   &                                &$-0.048$/$0.040$&$-0.050$/$0.039$&$\,\,\,\,\,\,\,0.000$/$0.000$ \\
  
\end{tabular}
\end{ruledtabular}
\end{table*}

\clearpage
\begin{table*}
\caption{Binding energies (in eV) (positive sign for states below the Fermi level and negative sign for states above the Fermi level) of the interface hybrid states ($I_i$) extracted from the calculated band structures of the graphene/FM system at the $\rm K$ and $\rm M$ points (see text for details).}\label{table:hybrid_states}
\begin{ruledtabular}
\begin{tabular}{ c  c  c  c  c } 
          &\multicolumn{2}{c}{FM = Ni(111)}&\multicolumn{2}{c}{FM = 1ML Fe/Ni(111)}\\
\cline{2-5}
\raisebox{1.5ex}[-1.5ex]{State} &spin $\uparrow$&spin $\downarrow$& spin $\uparrow$&spin $\downarrow$\\
\hline
$I_5$&$-3.20$&$-3.32$&$-3.56$&$-4.05$\\
$I_4$&$-0.12$&$-0.55$&$-0.28$&$-1.72$\\
$I_3$&$ 0.28$&$-0.16$&$  0.26$&$-1.19$\\
$I_2$&$ 2.28$&$ 1.93$&$  2.32$&$ 1.99$\\
$I_1$&$ 3.27$&$ 2.93$&$  3.75$&$ 2.77$\\
\end{tabular}
\end{ruledtabular}
\end{table*}

\clearpage
\begin{figure}
\includegraphics[scale=1.1]{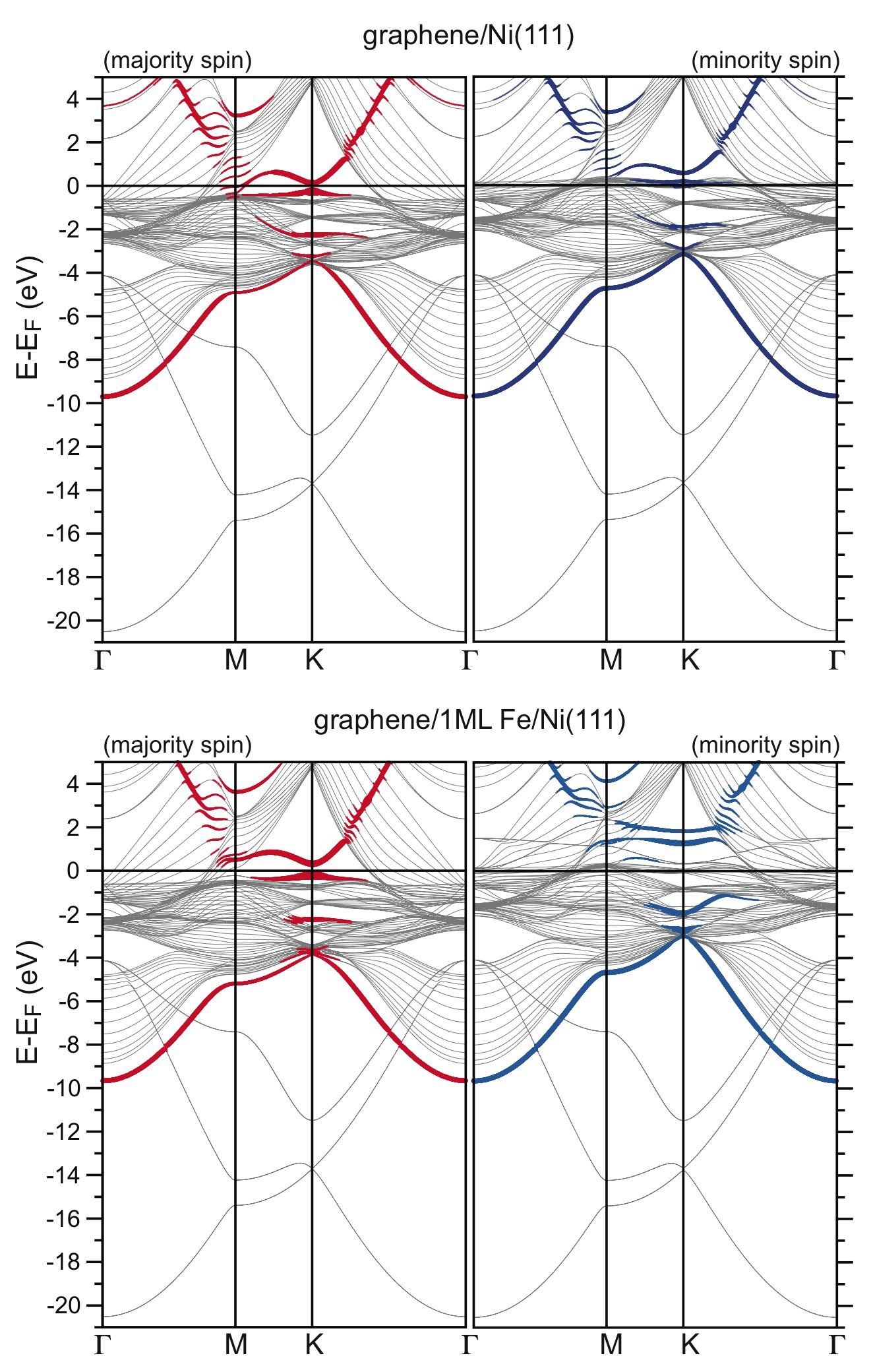}
\end{figure}
\textbf{Fig.\,S1.} Calculated majority and minority spin band structures of the graphene/Ni(111) and graphene/1\,ML\,Fe(111)/Ni(111) interfaces for a most energetically favorable configurations. For the blue/red (thicker) lines, the carbon $p_z$ character is used as a weighting factor.

\clearpage
\begin{figure}
\includegraphics[scale=0.85]{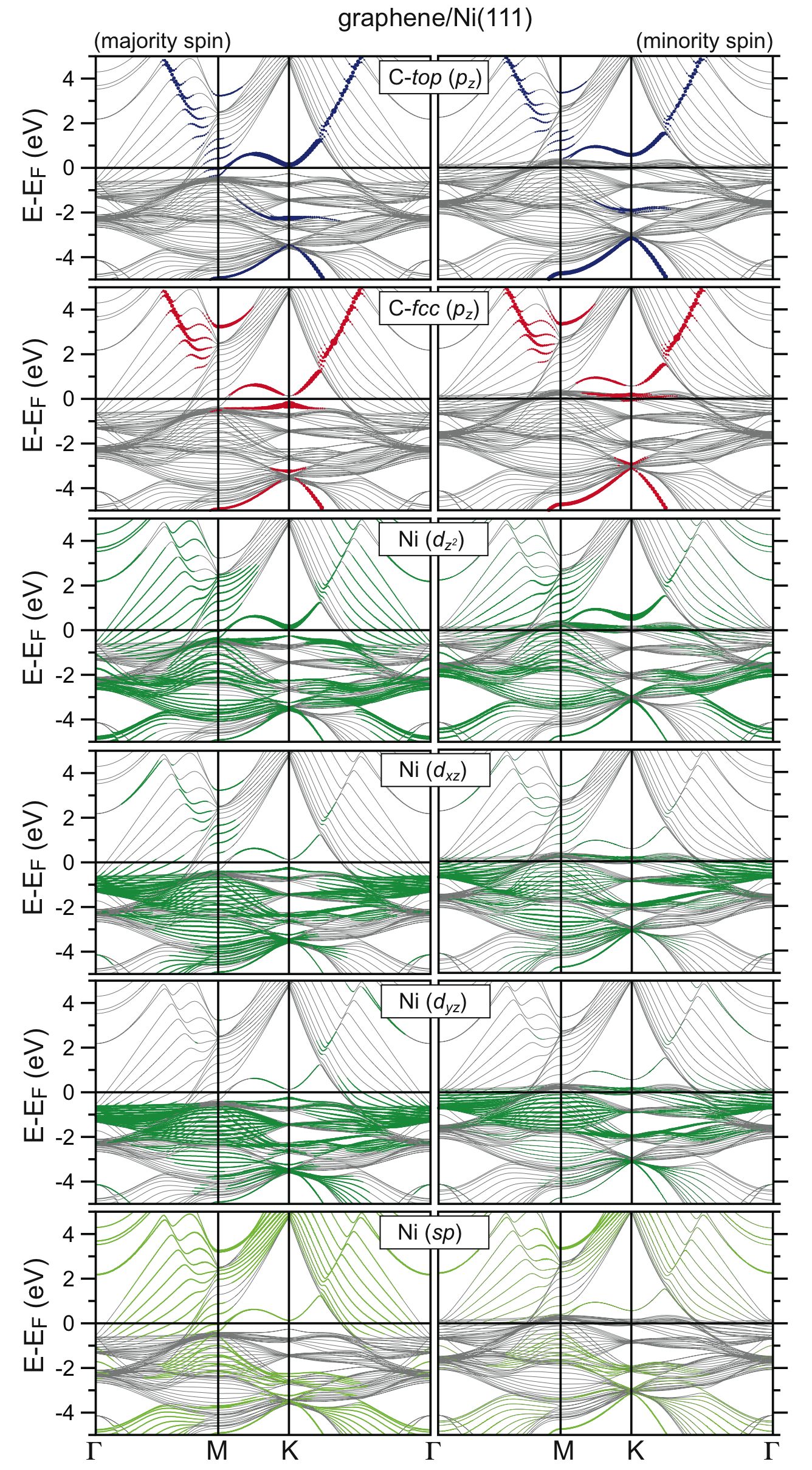}
\end{figure}
\textbf{Fig.\,S2.} Detailed analysis of the electronic structure of the graphene/Ni(111) system where the corresponding weights of Ni- and C-projected bands are shown by thick lines.

\clearpage
\begin{figure}
\includegraphics[scale=1.2]{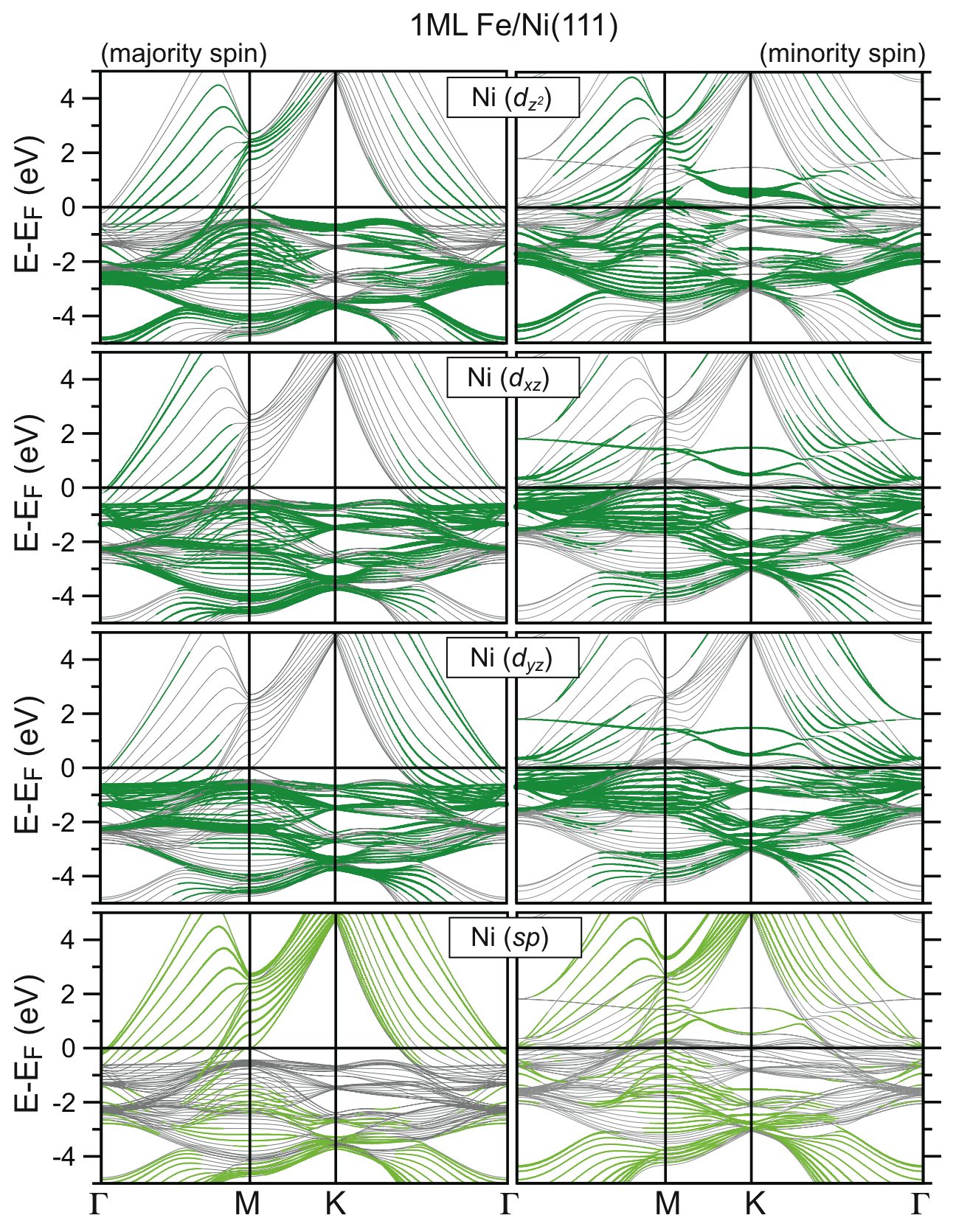}
\end{figure}
\textbf{Fig.\,S3.} Detailed analysis of the electronic structure of the 1\,ML\,Fe/Ni(111) system where the corresponding weights of Ni-projected bands are shown by thick lines.

\clearpage
\begin{figure}
\includegraphics[scale=1.2]{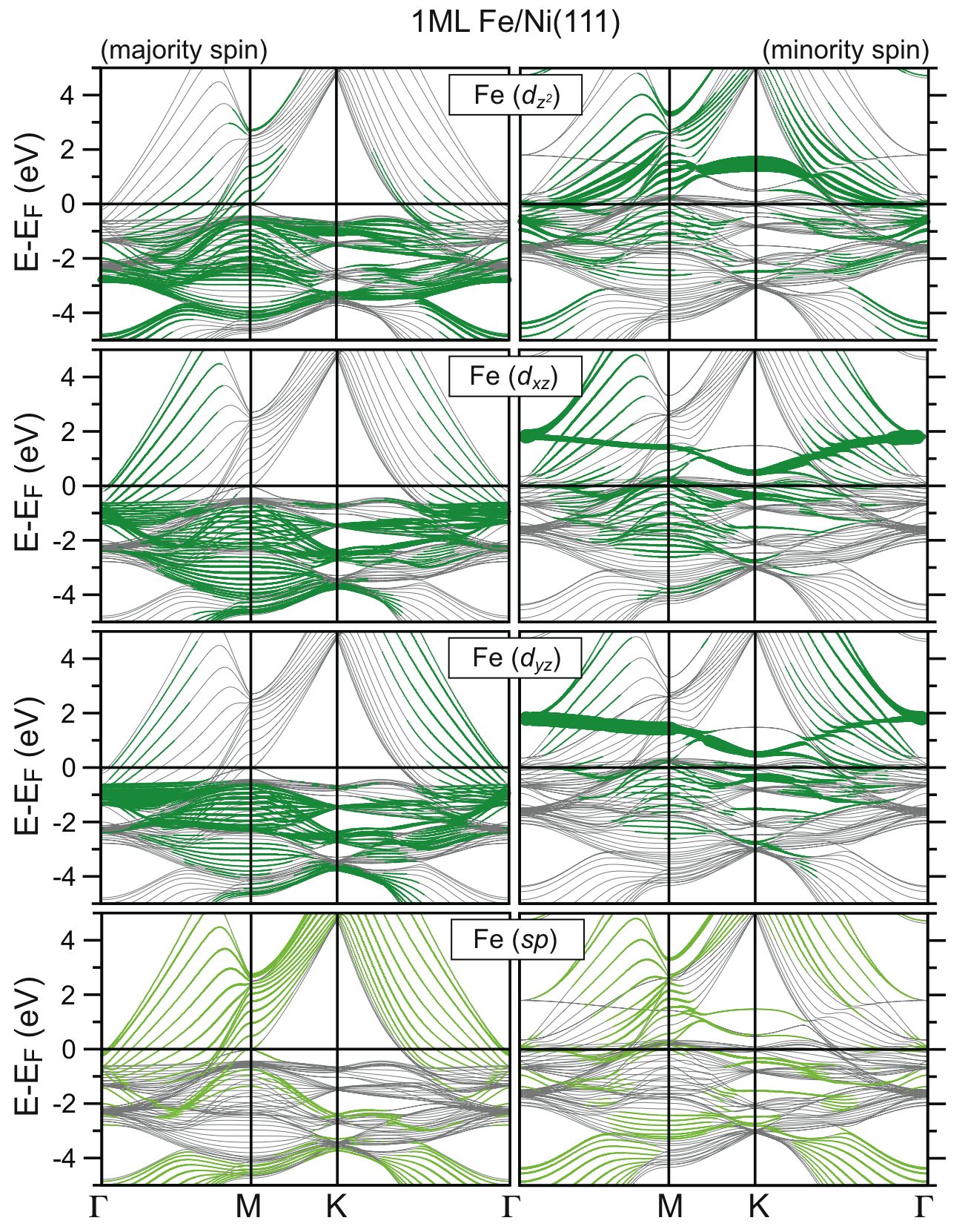}
\end{figure}
\textbf{Fig.\,S4.} Detailed analysis of the electronic structure of the 1\,ML\,Fe/Ni(111) system where the corresponding weights of Fe-projected bands are shown by thick lines.

\clearpage
\begin{figure}
\includegraphics[scale=0.85]{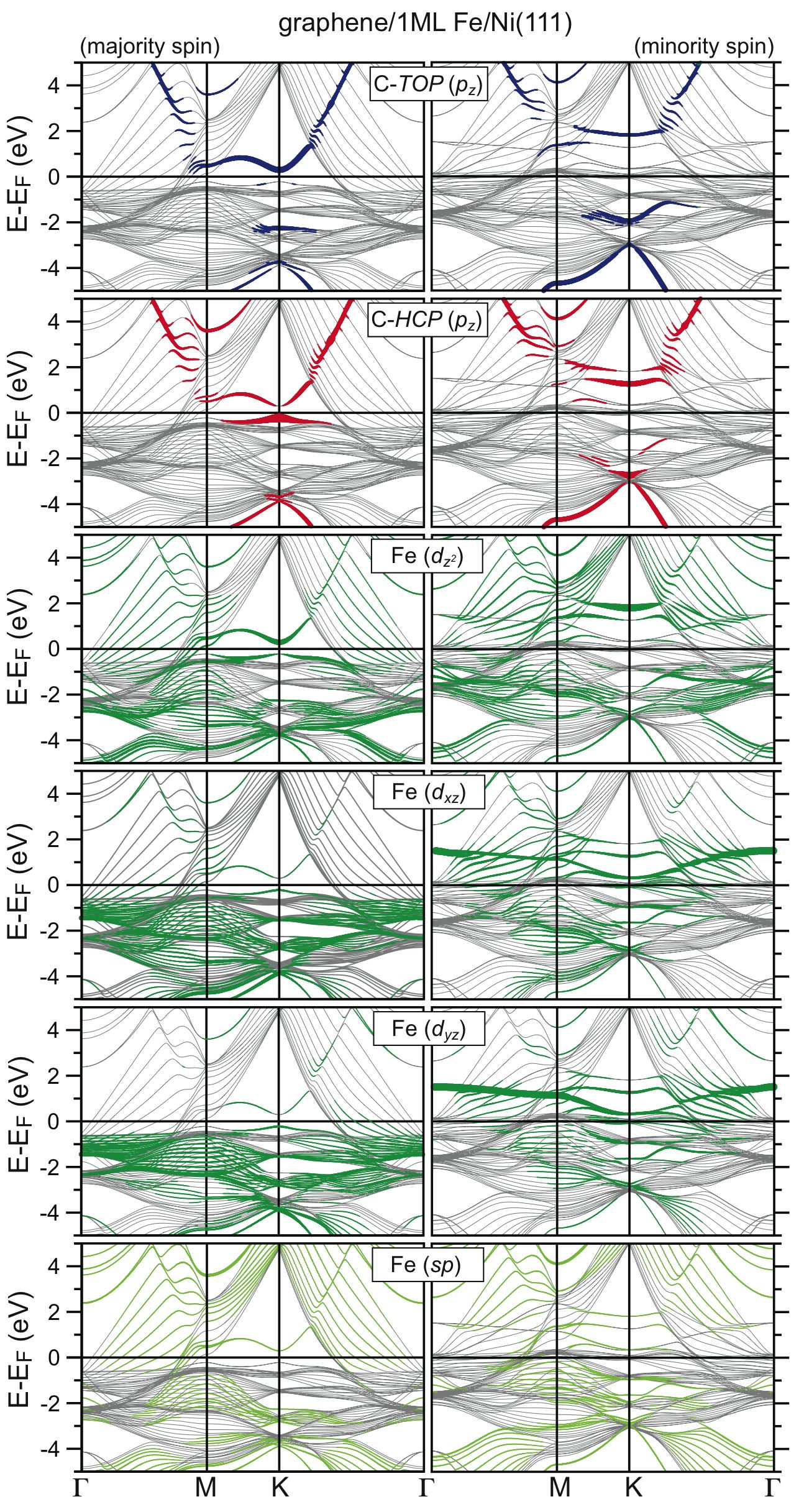}
\end{figure}
\textbf{Fig.\,S5.} Detailed analysis of the electronic structure of the graphene/1\,ML\,Fe(111)/ Ni(111) system where the corresponding weights of Fe- and C-projected bands are shown by thick lines.

\end{document}